# A Hybrid Residue–Floating Numerical Architecture with Formal Error Bounds for High-Throughput FPGA Computation

Mostafa Darvishi, *Senior Member, IEEE*

***Abstract*—Floating-point arithmetic is costly on FPGA platforms due to wide datapaths, normalization, and carry propagation, motivating alternative numerical representations that improve throughput and efficiency. This paper presents the** Hybrid Residue–Floating Numerical Architecture (HRFNA)**, a fully specified numerical system that combines carry-free residue arithmetic with lightweight exponent-based scaling to achieve wide dynamic range, predictable error behavior, and efficient FPGA implementation. HRFNA is developed with a rigorous mathematical foundation: the hybrid number space is formally defined, correctness of arithmetic and normalization is proven, and explicit absolute and relative error bounds are derived, confining rounding to infrequent normalization events. A complete FPGA microarchitecture is presented, featuring deeply pipelined modular arithmetic, exponent management, and a CRT-based normalization engine that sustains an initiation interval of one cycle under steady-state operation. Application-level evaluation on dot products, dense matrix multiplication, and iterative Runge–Kutta ODE solvers demonstrates stable numerical behavior over long computation sequences. Implemented on a Xilinx Zynq UltraScale+ ZCU104, HRFNA achieves up to 2.4× higher throughput, 38–55% LUT reduction, and up to 1.9× energy efficiency improvement compared to IEEE-754 FP32 baselines, while maintaining bounded numerical error. Comparative analysis shows that HRFNA occupies a previously unexplored design point between numerical stability, dynamic range, and hardware efficiency, making it well suited for FPGA-centric scientific and CAD-relevant computation.**

## I. INTRODUCTION

FIELD-programmable gate arrays (FPGAs) have become essential compute substrates for accelerating numerically intensive workloads in scientific computing, signal processing, control systems, and emerging heterogeneous platforms. Their ability to exploit massive spatial parallelism and fine-grained pipelining makes them particularly attractive for arithmetic-dominated kernels such as matrix operations, iterative solvers, and digital signal processing pipelines. In these domains, numerical representation plays a decisive role in determining performance, energy efficiency, and achievable accuracy. [1].

Mostafa Darvishi is with Electrical Engineering Department of École de technologie supérieure (ÉTS), Montreal, Canada. He is also VP of Engineering at Evolution Optiks R&D Inc. (e-mail: darvishi@ieee.org).

IEEE-754 floating-point arithmetic remains the dominant numerical standard due to its wide dynamic range, well-defined semantics, and software compatibility [2]. However, implementing floating-point arithmetic on FPGA fabrics is inherently expensive. Wide mantissa datapaths, exponent alignment, normalization, rounding logic, and multi-stage carry propagation significantly increase area, power consumption, and latency. Even with modern DSP-rich FPGA architectures and vendor-optimized floating-point IP cores, floating-point units often become throughput bottlenecks and limit the scalability of deeply pipelined designs [3].

To address these limitations, a wide spectrum of alternative numerical representations has been explored. Fixed-point arithmetic offers excellent hardware efficiency but lacks the dynamic range required for iterative refinement, multi-scale algorithms, and long accumulation chains [4], [5]. Logarithmic number systems (LNS) reduce multiplication to addition but incur substantial overhead for addition and subtraction due to costly antilogarithmic operations. Residue number systems (RNS) enable carry-free, highly parallel arithmetic that maps naturally onto FPGA fabrics, yet suffer from fundamental limitations in scaling, comparison, overflow detection, and fractional representation, often requiring expensive reconstruction through the Chinese Remainder Theorem (CRT) [6]-[9].

From a computer-aided design (CAD) perspective, these trade-offs reveal a critical gap. While FPGA designers can choose between floating-point generality and fixed-point efficiency, there is no widely adopted numerical abstraction that simultaneously provides: (i) carry-free, highly parallel arithmetic; (ii) wide and controllable dynamic range; (iii) bounded and analyzable numerical error; and (iv) hardware-efficient realization suitable for synthesis and timing closure.

Existing hybrid approaches attempt to mitigate individual shortcomings—for example, block floating-point systems reduce exponent overhead by sharing scale across vectors, and hybrid RNS–floating designs exploit residue arithmetic for large integer operations. However, most such systems are either domain-specific (e.g., cryptography or machine learning), lack a rigorous error model, or do not demonstrate stability and correctness at the application level. As a result, many proposed hybrid formats remain difficult to evaluate, compare, or integrate into general-purpose numerical pipelines.

This paper introduces the Hybrid Residue–Floating

Numerical Architecture (HRFNA), a numerical system designed to explicitly bridge this gap. HRFNA unifies carry-free residue arithmetic with a compact exponent-based scaling mechanism, decoupling integer arithmetic from dynamic-range management. In HRFNA, arithmetic operations are performed predominantly in the residue domain, preserving full parallelism and eliminating carry propagation, while a lightweight exponent tracks global scaling. Unlike prior residue-based or hybrid approaches, HRFNA is developed as a fully specified numerical system, not merely an architectural optimization.

The central design philosophy of HRFNA is that *normalization and rounding should be rare, structured, and analyzable events*, rather than pervasive operations embedded in every datapath stage. To this end, HRFNA introduces deterministic CRT-based normalization governed by explicit thresholds, allowing reconstruction-driven scaling to occur infrequently while guaranteeing bounded numerical error. This approach enables long-running iterative computations—such as dot-product accumulation and differential equation solvers—to execute stably without the continuous overhead of floating-point normalization.

The contributions of this work are summarized as follows: (i) *Formal Numerical Model*: We define a hybrid residue–floating number space and its semantic mapping, establishing a precise mathematical foundation for HRFNA; (ii) *Correctness and Error Analysis*: We prove correctness of hybrid multiplication and normalization and derive explicit absolute and relative error bounds, showing that HRFNA behaves as a deterministic block-floating–like system with carry-free arithmetic; (iii) *FPGA Microarchitecture*: We present a deeply pipelined FPGA architecture that sustains an initiation interval of one cycle under steady-state operation, integrating modular arithmetic, exponent management, and CRT-based normalization; (iv) *Application-Level Validation*: We evaluate HRFNA on canonical numerical workloads, including dot products, matrix multiplication, and Runge–Kutta ODE solvers, demonstrating numerical stability over long execution sequences; and (v) *Comparative Evaluation*: We provide comprehensive comparison against IEEE-754 floating-point, fixed-point, block floating-point, and prior hybrid RNS-based systems, clarifying the novelty and practical impact of HRFNA.

By combining mathematical rigor, architectural efficiency, and application-level validation, HRFNA establishes a new design point between floating-point generality and fixed-point efficiency. The remainder of this paper develops the theoretical foundations, algorithms, and hardware realization of HRFNA and demonstrates its suitability for FPGA-centric numerical computation.

## II. Background and Related Works

The design of numerical representations for FPGA-based computation has been an active research area for decades, driven by the need to balance dynamic range**,** numerical accuracy**, and** hardware efficiency [3], [5], [8]. This section reviews prior work across four major classes of numerical systems relevant to FPGA-centric computation and positions HRFNA with respect to their capabilities and limitations [2], [4]. The discussion emphasizes both numerical properties and architectural implications, reflecting the dual algorithm–architecture focus expected in CAD-oriented research [9]-[13].

### *A. Floating-Point Arithmetic on FPGAs*

IEEE-754 floating-point arithmetic provides a standardized and widely adopted numerical format with well-defined semantics, rounding behavior, and exceptional-case handling. As a result, floating-point remains the default choice for general-purpose numerical computation. However, its implementation on FPGA fabrics is intrinsically costly. Floating-point addition and multiplication require exponent comparison, alignment, normalization, rounding, and multi-stage carry propagation, resulting in wide datapaths and deep pipelines [10], [11].

Substantial prior work has focused on optimizing floating-point units for FPGA deployment, including reduced-precision formats, fused multiply–add units, and DSP48-based implementations. While these efforts have improved frequency and energy efficiency, they do not fundamentally eliminate the structural overhead of floating-point normalization. Even vendor-provided floating-point IP cores remain resource-intensive, particularly when instantiated at scale or deeply pipelined. Consequently, floating-point arithmetic often becomes a throughput and area bottleneck in FPGA accelerators dominated by arithmetic kernels [12]- [17].

### *B. Fixed-Point Arithmetic and Its Limitations*

Fixed-point arithmetic offers a compelling alternative when numerical range and precision requirements are known a priori. Its hardware efficiency stems from narrow datapaths, simple control logic, and the absence of normalization and rounding circuitry. Fixed-point arithmetic is therefore widely used in signal processing and machine learning inference [13].

However, fixed-point representations lack the dynamic range required for many scientific and control applications, particularly those involving iterative refinement, accumulation over long sequences, or multi-scale phenomena [11], [14]. Preventing overflow or underflow typically requires conservative scaling choices, which in turn reduce effective precision. While mixed-precision and adaptive scaling techniques have been proposed, they often reintroduce complexity comparable to floating-point arithmetic without providing the same generality [12]-[17] .

### *C. Logarithmic Number Systems*

Logarithmic number systems (LNS) represent values in the logarithmic domain, transforming multiplication and division into addition and subtraction. This property makes LNS attractive for multiplication-heavy workloads. However, addition and subtraction in LNS require conversion back to the linear domain through logarithm and antilogarithm operations, which are expensive and sensitive to approximation error [18].

Table I. Qualitative Comparison of Numerical Representations

| Representation | Carry-Free | Dynamic Range | Formal Error Model | FPGA-Validated | Application-Level Stability |
|---|---|---|---|---|---|
| Fixed-Point | ✗ | ✗ | ✓ | ✓ | Limited |
| IEEE-754 FP | ✗ | ✓ | ✓ | ✓ | ✓ |
| Block FP | ✗ | ✓ | Partial | ✓ | Limited |
| Pure RNS | ✓ | ✗ | ✗ | ✓ | ✗ |
| Prior Hybrid RNS | ✓ | Partial | ✗ | Partial | ✗ |
| **HRFNA** | ✓ | ✓ | ✓ | ✓ | ✓ |

Several hybrid LNS-based architectures have been proposed to mitigate these costs, often combining logarithmic multiplication with linear-domain addition. While such designs can reduce multiplication latency, they typically suffer from high control complexity, limited numerical robustness, and poor scalability for general-purpose computation. As a result, LNS remains a niche solution rather than a broadly applicable numerical abstraction for FPGA-based computation [14]-[19].

### *D. Residue Number Systems*

Residue number systems (RNS) represent integers as vectors of residues with respect to a set of pairwise coprime moduli. A defining advantage of RNS is that arithmetic operations such as addition and multiplication can be performed independently in each residue channel, completely eliminating carry propagation. This property makes RNS highly attractive for FPGA fabrics, which naturally support parallel, bit-sliced computation [18]- [20].

RNS has been successfully applied in cryptography, digital signal processing, and large-integer arithmetic. However, pure RNS suffers from fundamental limitations that hinder its use in general numerical computation. Comparison, sign detection, scaling, and overflow detection are nontrivial and often require reconstruction via the Chinese Remainder Theorem (CRT), which is expensive in both area and latency. Moreover, representing fractional values and maintaining numerical stability over long computation sequences remain open challenges in pure RNS systems [20], [21].

Numerous extensions to RNS have been proposed, including mixed-radix representations, scaling-friendly RNS variants, and approximate comparison techniques. While these approaches reduce reconstruction overhead, they often introduce significant control complexity, reduce parallelism, or require restrictive modulus choices, limiting their practicality in high-throughput FPGA pipelines [19], [21].

### *E. Block Floating-Point and Shared-Exponent Systems*

Block floating-point (BFP) and shared-exponent systems represent vectors or blocks of values using a common exponent and individual fixed-point mantissas. These formats reduce exponent handling overhead and are widely used in signal processing and machine learning accelerators. BFP provides improved dynamic range relative to fixed-point arithmetic while retaining much of its hardware efficiency [22]. However, BFP systems require careful management of block boundaries and scaling decisions. When values within a block diverge significantly in magnitude, precision loss becomes unavoidable. Furthermore, BFP does not eliminate carry propagation within mantissa arithmetic and therefore cannot fully exploit the parallelism available in FPGA fabrics [23].

### *F. Hybrid Numerical Systems*

Hybrid numerical systems combine elements of multiple representations to exploit complementary advantages. Prior work has explored hybrid RNS–floating architectures for cryptographic accelerators, where residue arithmetic handles large integer operations while floating-point logic manages scaling. Other hybrid approaches combine logarithmic multiplication with linear-domain accumulation or employ shared-exponent schemes across SIMD lanes [23]-[25].

Despite their promise, most existing hybrid systems are *domain-specific* and lack a unified numerical model suitable for general-purpose computation. In particular, many do not provide: (i) a formal definition of the numerical representation; (ii) provable correctness guarantees; (iii) explicit error bounds associated with scaling or normalization, or (iv) application-level validation demonstrating long-term numerical stability. As a result, it is often difficult to assess their novelty, correctness, or suitability for integration into CAD-driven design flows [26], [27].

### *G. Positioning of HRFNA*

The Hybrid Residue–Floating Numerical Architecture (HRFNA) introduced in this work is designed to address the limitations identified above [28]-[31]. HRFNA combines carry-free residue arithmetic with exponent-based scaling in a manner that preserves parallelism while enabling dynamic-range management. Unlike prior hybrid approaches, HRFNA is developed as a *fully specified numerical system* with a rigorous mathematical foundation, bounded error behavior, and a hardware architecture validated at both microarchitectural and application levels. Table I summarizes the qualitative differences between representative numerical systems and highlights the design gap that HRFNA fills. This comparison clarifies that HRFNA occupies a previously unexplored design point that simultaneously delivers carry-free arithmetic, wide dynamic range, analyzable error behavior, and demonstrated numerical stability, making it well suited for FPGA-centric numerical computation.

## III. NUMERICAL REPRESENTATION AND HRFNA FOUNDATION

The Hybrid Residue–Floating Numerical Architecture (HRFNA) is designed as a *numerical system*, not merely a hardware optimization. Consequently, its development must begin with a precise mathematical formulation that defines the representable number space, establishes correctness of arithmetic operations, and characterizes the error behavior introduced by scaling and normalization. This section provides that foundation.

The formulation intentionally separates *carry-free integer arithmetic* from dynamic-range management, enabling parallel residue-domain computation while retaining controlled scaling behavior comparable to block floating-point systems. This separation is central to HRFNA's ability to combine numerical rigor with FPGA-efficient implementation [32], [33].

### *A. Hybrid Residue–Floating Number Space*

Let $\{m_1, m_2, \ldots, m_k\}$ be a set of pairwise coprime moduli, and let

$$M = \prod_{i=1}^{k} m_i$$

denote the composite modulus. In classical residue number systems (RNS), any integer $N \in [0, M)$ is uniquely represented by its residue vector

$$\boldsymbol{r} = (r_1, r_2, \cdots, r_k), \qquad r_i = N \ mod \ m_i$$

HRFNA extends this representation by associating each residue vector with a **global exponent** that governs numerical scale.

**Definition 1 (HRFNA Number Space)**
The HRFNA number space is defined as

$$\mathcal{H} = \{(\boldsymbol{r}, f) \mid \boldsymbol{r} \in Z_{m_1} \times \cdots \times Z_{m_k}, \quad f \in Z\}$$

Each hybrid number $(\boldsymbol{r}, f)$ maps to a real value through the semantic function

$$\Phi(\boldsymbol{r}, f) = CRT(\boldsymbol{r}) \cdot 2^f$$

where $CRT(\cdot)$ denotes reconstruction via the Chinese Remainder Theorem. This formulation explicitly decouples *integer magnitude representation* (handled entirely in the residue domain) from scaling (handled by the exponent). Fractional values are represented implicitly through negative exponents, while residue arithmetic remains strictly integer.

**Proposition 1 (Uniqueness of Representation)**
For a fixed modulus set $\{m_i\}$ and exponent $f$, the mapping $\Phi: \mathcal{H} \to \mathbb{R}$ is injective for all reconstructed integers in the interval $[0, M)$.

**Proof:**
The Chinese Remainder Theorem guarantees a unique integer reconstruction in $[0, M)$ for each residue vector $\mathbf{r}$. Multiplication by a fixed scaling factor $2^f$ preserves injectivity.

### *B. Hybrid Multiplication*

A primary design goal of HRFNA is to preserve the *full parallelism* of residue arithmetic during multiplication, while handling scaling through a lightweight operation. Let two hybrid numbers be given by

$$X = (\boldsymbol{r}_X, f_X), \quad Y = (\boldsymbol{r}_Y, f_Y)$$

**Definition 2 (Hybrid Multiplication)**
Hybrid multiplication is defined as

$$Z = X \otimes Y = (\boldsymbol{r}_Z, f_Z)$$

where

$$\boldsymbol{r}_Z = \boldsymbol{r}_X \odot \boldsymbol{r}_Y, \quad f_Z = f_X + f_Y$$

and $\odot$ denotes element-wise residue multiplication:

$$r_{Z,i} = (r_{X,i} \cdot r_{Y,i}) \ mod \ m_i$$

**Theorem 1 (Correctness of Hybrid Multiplication)**
For all $X, Y \in \mathcal{H}$ such that normalization is not triggered,

$$\Phi(X \otimes Y) = \Phi(X) \cdot \Phi(Y)$$

**Proof:**
By the homomorphic property of CRT,

$$CRT(\boldsymbol{r}_X \odot \boldsymbol{r}_Y) = CRT(\boldsymbol{r}_X) \cdot CRT(\boldsymbol{r}_Y)$$

Exponent addition yields

$$CRT(\boldsymbol{r}_X \odot \boldsymbol{r}_Y) \cdot 2^{f_X + f_Y} = (CRT(\boldsymbol{r}_X) \cdot 2^{f_X})(CRT(\boldsymbol{r}_Y) \cdot 2^{f_Y})$$

Thus, $\Phi(Z) = \Phi(X)\Phi(Y)$.

This result establishes that HRFNA multiplication is *exact* prior to normalization and requires *no carry propagation, no exponent alignment, and no rounding* during the core arithmetic operation.

### *C. Dynamic Range and Normalization*

Repeated multiplication and accumulation in the residue domain can cause the reconstructed integer magnitude to approach or exceed the representable range $[0, M)$. HRFNA addresses this through *explicit, threshold-based normalization*.

Let $N = \mathrm{CRT}(\mathbf{r})$ denote the reconstructed integer associated with a hybrid number.

**Definition 3 (Normalization Threshold)**
Let $\tau$ be a predefined threshold such that

$$\tau < M$$

Normalization is triggered when $| N | \geq \tau$. The threshold $\tau$ is chosen to provide sufficient headroom for continued residue arithmetic while ensuring that normalization events remain infrequent.

**Definition 4 (Hybrid Normalization)**

When normalization is triggered, the reconstructed integer $N$ is scaled by a power-of-two factor

$$K = 2^s$$

yielding

$$\tilde{N} = \left\lfloor \frac{N}{K} \right\rfloor, \quad \tilde{f} = f + s$$

The value $\tilde{N}$ is then re-encoded into the residue domain, producing a new hybrid number $(\tilde{\mathbf{r}}, \tilde{f})$.

*D. Error Analysis*

Normalization is the *only source of numerical error* in HRFNA. Importantly, this error is *deterministic, bounded, and analyzable*.

**Lemma 1 (Absolute Error Bound)**

Let $X = (\mathbf{r}, f)$ be normalized using scale factor $K = 2^s$. The absolute error introduced by normalization satisfies

$$|\epsilon| \leq 2^{f+s-1}$$

**Proof:**

Rounding via floor division introduces at most half of the scaling unit in reconstructed value space. Scaling by $2^{f+s}$ yields the stated bound.

**Lemma 2 (Relative Error Bound)**

The relative error introduced by normalization satisfies

$$\frac{|\epsilon|}{|\Phi(X)|} \leq 2^{-s}$$

**Proof:**

Immediate from Lemma 1 and the definition of normalization scaling.

**Interpretation**

These bounds show that HRFNA behaves as a *deterministic block-floating-point system*, with two key distinctions: (i) *Rounding occurs only during normalization*, not during every arithmetic operation; and (ii) *Core arithmetic remains carry-free and exact* within the residue domain. This combination enables long-running iterative computations with predictable error growth and reduced hardware overhead.

*E. Conceptual Interpretation and Magnitude Management in HRFNA*

While the formal development in Sections III-A through III-D establishes HRFNA as a mathematically well-defined numerical system, it is equally important to illustrate how magnitude, scaling, and comparison are handled in practice without pervasive reconstruction or carry propagation. Fig. 1 provides a conceptual overview of these mechanisms and clarifies how HRFNA differs fundamentally from both conventional floating-point arithmetic and pure residue number systems. The left-side of Fig. 1(a) illustrates an array of HRFNA values represented purely in the residue domain. Each value is stored as a vector of residues

$$(x_1, x_2, \cdots, x_k)$$

corresponding to the moduli $\{m_1, \dots, m_k\}$. At this stage, no reconstruction or binary conversion is performed, preserving the carry-free and fully parallel nature of residue arithmetic. To enable magnitude-aware operations—such as normalization decisions, comparison, and selection—each residue vector is augmented with a *floating-point interval evaluation* $[f, f]$, computed using lightweight interval arithmetic techniques. These intervals provide a conservative estimate of the reconstructed magnitude without explicitly computing $\mathrm{CRT}(\mathbf{r})$. Crucially, this interval evaluation: (i) avoids full CRT reconstruction, (ii) incurs significantly lower hardware cost than exact magnitude recovery, and (iii) and is sufficient to guide normalization and comparison decisions. Each interval is associated with an index (`idx`) identifying its corresponding residue vector, enabling deferred selection without loss of correspondence between magnitude estimates and residue-domain data.

The right-side of Fig. 1(a) shows a reduction tree that operates exclusively on floating-point interval evaluations rather than reconstructed integers. Pairwise comparisons are performed hierarchically to identify the residue representation with the largest estimated magnitude. Each node in the reduction tree propagates: (i) the selected interval $[f, f]$, and (ii) the corresponding index (`idx`). This structure enables efficient magnitude comparison across large arrays of HRFNA values using: (a) standard floating-point comparators, (b) logarithmic depth, and (c) without disturbing the residue-domain data. At the conclusion of the reduction, the index of the dominant magnitude is used to *select the corresponding residue vector* (denoted $X^{\backslash *}$) for normalization or scaling decisions. Importantly, this process does *not* require reconstruction of all candidates—only the selected element may be reconstructed if normalization is triggered. This mechanism directly supports the deterministic, threshold-driven normalization model developed in Section III-C.

Fig. 1(b) contrasts HRFNA's magnitude management strategy with conventional binary and IEEE-754 floating-point representations. In standard binary arithmetic, magnitude is implicit in bit position, and every arithmetic operation is subject to carry propagation. IEEE-754 floating-point

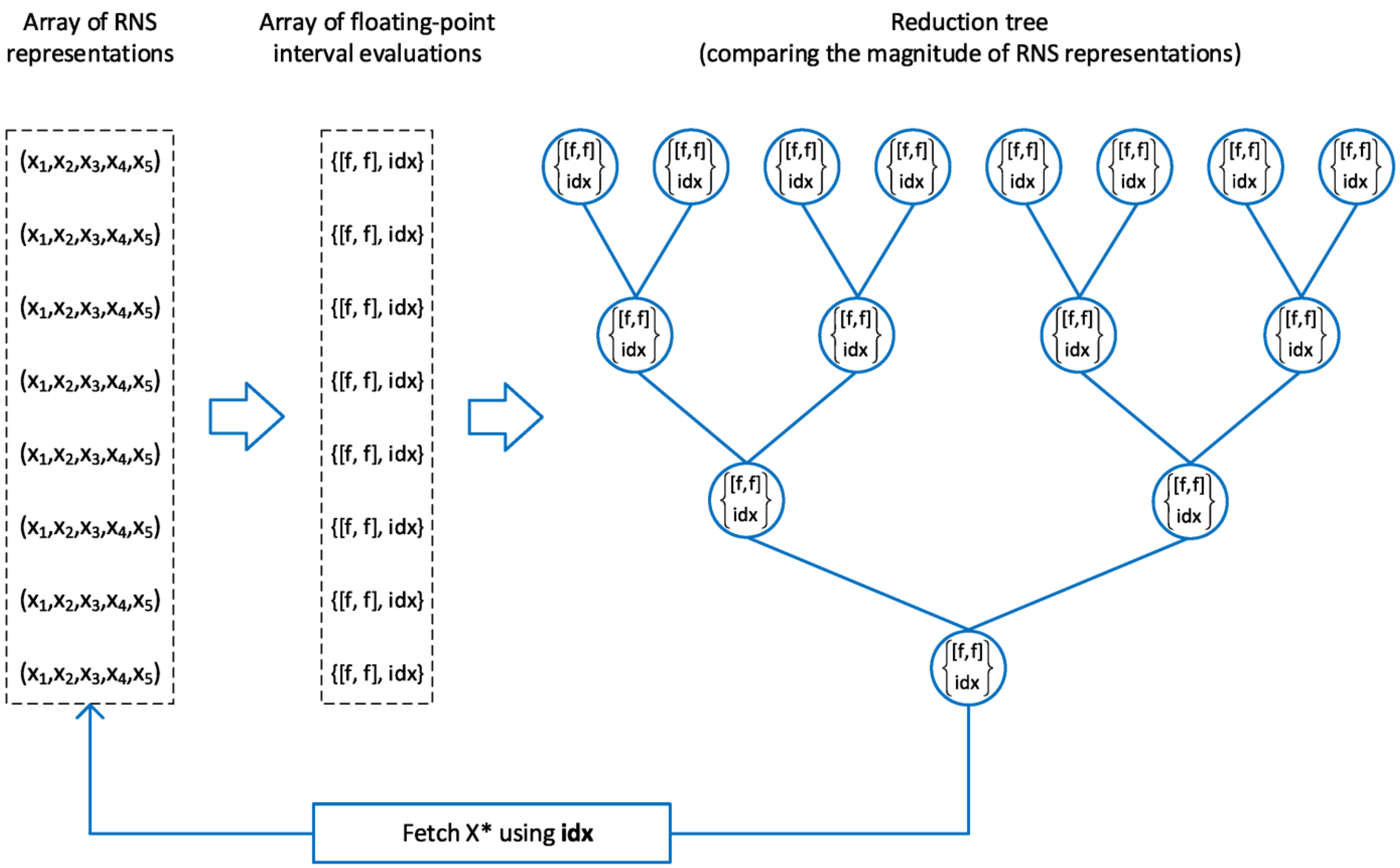

**(a)**

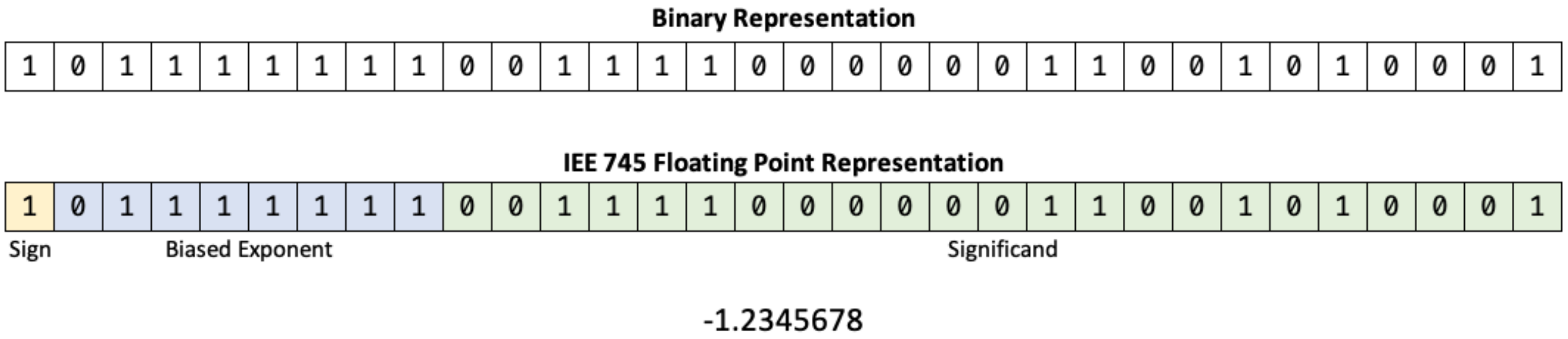

**(b)**

Fig. 1. Conceptual illustration of HRFNA representation. Integer arithmetic is performed in parallel residue channels, while a single global exponent controls scaling. Normalization is triggered only when residue-domain magnitude exceeds a threshold.

mitigates dynamic-range limitations through explicit exponent fields but requires exponent alignment, normalization, and rounding at nearly every operation. In contrast, HRFNA: (i) performs *core arithmetic entirely in the residue domain*, free of carry propagation, (ii) tracks scale using a *global exponent*, updated only when normalization occurs, and (iii) relies on *interval-based magnitude estimation* rather than exact reconstruction for most control decisions. As a result, normalization and rounding are *explicit, infrequent, and analyzable events*, rather than ubiquitous features of the datapath. This distinction underpins HRFNA's favorable balance between numerical rigor and hardware efficiency.

Together, Fig. 1(a),(b) illustrates the central design principle of HRFNA, i.e., *separating magnitude management from arithmetic execution*. From a numerical standpoint, this separation ensures that: (a) exact arithmetic is preserved between normalization events, (b) error is introduced only during controlled scaling operations, and (c) error growth is bounded as shown in Section III-D. From an architectural standpoint, it enables: (i) parallel, latency-matched residue pipelines, (ii) predictable control flow driven by interval comparison, and (iii) synthesis-friendly hardware with bounded normalization frequency. This conceptual framework directly supports the FPGA microarchitecture developed in Section V and the application-level stability demonstrated in Section VII.

## IV. HYBRID ARITHMETIC ALGORITHMS

The mathematical formulation of HRFNA in Section III defines the numerical representation, correctness guarantees, and bounded error behavior. This section translates those foundations into *explicit arithmetic algorithms* suitable for hardware realization. The focus is on structuring computation such that carry-free residue arithmetic is maximally exploited, while normalization and scaling are invoked only when required by magnitude growth. We first describe primitive arithmetic operations and then present composite algorithms commonly used in scientific and CAD-relevant workloads.

### *A. Hybrid Multiplication*

Hybrid multiplication is the fundamental operation in HRFNA and directly reflects the formulation in Section III-B. Given two hybrid numbers

$$X = (\boldsymbol{r}_X, f_X), \quad Y = (\boldsymbol{r}_Y, f_Y)$$

multiplication proceeds as follows:
(i) Residue-domain multiplication: Each residue channel computes

$$r_{Z,i} = (r_{X,i} \cdot r_{Y,i}) \ mod \ m_i$$

independently and in parallel.

(ii) Exponent update: The global exponent is updated as

$$f_Z = f_X + f_Y$$

(iii) Magnitude check (optional): Interval evaluation logic (Section III-E) determines whether normalization is required. No exponent alignment, rounding, or carry propagation is involved in the core multiplication step. As proven in Theorem 1, the operation is *exact prior to normalization*, making it well suited to deep pipelining and high-throughput execution.

### *B. Hybrid Addition and Exponent Synchronization*

Addition in HRFNA differs from multiplication in that it requires operands to share a common exponent. Let

$$X = (\boldsymbol{r}_X, f_X), \quad Y = (\boldsymbol{r}_Y, f_Y)$$

If $f_X \neq f_Y$, one operand must be rescaled to match the other. HRFNA enforces *explicit exponent synchronization*, avoiding implicit alignment logic in the datapath. The exponent synchronization strategy is performed as follows: Without loss of generality, assume $f_X > f_Y$. Then: (i) Compute exponent difference $\Delta = f_X - f_Y$; (ii) Scale $Y$ by $2^{-\Delta}$ using controlled normalization; and finally (iii) Update $Y$ to $(\tilde{\mathbf{r}}_{Y'} f_X)$. Once synchronized, residue-domain addition proceeds as:

$$r_{Z,i} = (r_{X,i} \cdot r_{Y,i}) \ mod \ m_i \quad f_Z = f_X$$

Because exponent synchronization is explicit and infrequent, HRFNA avoids the continuous exponent alignment overhead present in IEEE-754 floating-point addition.

### *C. Hybrid Multiply-Accumulate (MAC)*

Multiply–accumulate operations are central to dot products, matrix multiplication, and iterative solvers. HRFNA supports MAC through a structured sequence of hybrid multiplications followed by synchronized accumulation. Given accumulator $A = (\mathbf{r}_A, f_A)$ and operands $X$ and $Y$: (i) Compute product $P = X \otimes Y$; (ii) Synchronize exponents of $A$ and $P$; and (iii) Accumulate residues:

$$\boldsymbol{r}_A \leftarrow \boldsymbol{r}_A + \boldsymbol{r}_P \quad (mod \ m_i)$$

Normalization is deferred until a magnitude threshold is reached, allowing long accumulation sequences without reconstruction.

### *D. Hybrid Dot-Product Algorithm*

Dot products are representative of accumulation-heavy workloads and are particularly sensitive to rounding error in floating-point arithmetic. HRFNA addresses this through *exponent-coherent accumulation*.

**Algorithm 1: Hybrid Dot Product**
**Input**:

$\{(X_j, Y_j)\}_{j=1}^{N}$ where

$X_j = \boldsymbol{r}_{X_j}, f_{X_j}$

$Y_j = \boldsymbol{r}_{Y_j}, f_{Y_j}$

**Output**:

Dot product $\boldsymbol{S}$

**Steps**:

1. Initialize accumulator $A = (\mathbf{0}, f_0)$, where $f_0$ is chosen to match initial operands.
2. For $j = 1$ to $N$:
   a. Compute $P_j = X_j \otimes Y_j$
   b. Synchronize exponent of $P_j$ with $A$
   c. Accumulate residues: $\mathbf{r}_A \leftarrow \mathbf{r}_A + \mathbf{r}_{P_j}$
3. Periodically check magnitude using interval evaluation.
4. If $|A| \geq \tau$, apply normalization.
5. After final iteration, reconstruct $A$ once to obtain the result.

**Numerical Implications**:
This algorithm ensures that: (i) rounding occurs only during normalization, (ii) accumulation error grows predictably, and (iii) long dot products remain stable, as confirmed in Section VII.

### E. Matrix Multiplication and Kernel Composition

Matrix multiplication is constructed by composing hybrid dot products. Because HRFNA maintains exponent coherence across accumulation, inner products can be computed efficiently without per-element exponent realignment. For an $N \times N$ matrix multiplication: (i) each output element invokes one Hybrid Dot Product, (ii) normalization frequency depends on operand magnitudes rather than matrix size, and (iii) and residue arithmetic remains fully parallel across channels. This composability demonstrates that HRFNA supports *structured numerical kernels*, not just isolated operations.

### F. Algorithmic Advantages Over Conventional Floating-Point

Compared with IEEE-754 floating-point arithmetic, HRFNA algorithms exhibit: (i) *Reduced normalization frequency*: normalization is threshold-driven rather than operation-driven; (ii) *Exact intermediate arithmetic*: multiplication and addition are exact in the residue domain; and (iii) *Predictable error growth*: error bounds follow directly from normalization parameters. These properties are critical for CAD-relevant numerical workloads, where deterministic behavior and analyzable error are often more important than strict IEEE compliance.

## V. HARDWARE MICROARCHITECTURE

The Hybrid Residue–Floating Numerical Architecture (HRFNA) is designed to translate the mathematical and algorithmic structure developed in Sections III and IV into a high-throughput, synthesis-friendly FPGA implementation. The central architectural objective is to preserve carry-free, fully parallel arithmetic in the residue domain while ensuring that scaling and normalization remain explicit, bounded, and decoupled from the critical datapath.

To achieve this, the HRFNA microarchitecture is organized into three loosely coupled subsystems: (i) a residue arithmetic pipeline, (ii) an exponent management pipeline, and (iii) a CRT-based normalization engine. This separation mirrors the conceptual structure illustrated in Fig. 1 and enables predictable control flow and scalable performance.

Fig. 2 presents the top-level datapath organization of HRFNA. Hybrid numbers enter the architecture as pairs $(\boldsymbol{r}, f)$, where the residue vector $\mathbf{r}$and exponent $f$are routed through *logically independent pipelines*. The residue components are dispatched to an array of parallel modular arithmetic units, one per modulus, implementing the carry-free operations defined in Section IV. In parallel, the exponent values are routed through a lightweight integer pipeline responsible for exponent updates and synchronization. This explicit separation of residue-domain arithmetic from exponent management directly reflects the formal numerical model introduced in Section III and is central to HRFNA's scalability. Importantly, Fig. 2 highlights that *no normalization or reconstruction logic lies on the critical arithmetic path*, enabling uninterrupted pipelined execution under steady-state conditions.

Fig. 3 focuses on the control path responsible for magnitude tracking and normalization decisions. Rather than reconstructing residue values at every operation, HRFNA employs the interval-evaluation mechanism described in Section III-E to estimate magnitude conservatively.

In this subfigure, floating-point interval evaluations are generated alongside arithmetic results and fed into a comparison network that checks against a predefined threshold. When the estimated magnitude exceeds the threshold, a normalization request is issued. Otherwise, arithmetic proceeds without interruption.

This structure ensures that magnitude management is *decoupled from arithmetic execution*, allowing normalization to be both infrequent and deterministic. Fig. 2(b) thus provides the architectural realization of the error-bounded normalization model developed in Section III-D.

Fig. 4 isolates the CRT-based normalization engine, which is activated only when explicitly requested by the control path in Fig. 2(b). Upon activation, the selected residue vector is reconstructed using the Chinese Remainder Theorem, scaled by a power-of-two factor, and re-encoded into the residue domain. The global exponent is updated accordingly. By placing the normalization engine off the main datapath, HRFNA ensures that the latency associated with reconstruction and scaling does not affect the steady-state throughput of arithmetic pipelines. This architectural choice directly supports the throughput guarantee proven in Theorem 2 and enables normalization latency to be amortized over many arithmetic operations.

Together, Fig. 2 to Fig. 4 illustrate the central architectural principle of HRFNA: *structured separation of arithmetic, control, and normalization*. Residue-domain computation remains fully parallel and carry-free, exponent handling is lightweight and explicit, and normalization is invoked only when required by magnitude growth. This decomposition not only improves performance and energy efficiency but also simplifies synthesis, timing closure, and design-space exploration—key considerations for CAD-driven FPGA design.

## VI. RTL IMPLEMENTATION

This section describes the register-transfer level (RTL) realization of HRFNA and the implementation choices that enable a synthesizable, timing-closed design on modern FPGA fabrics. The RTL design follows the architectural partitioning introduced in Section V: (i) residue arithmetic pipelines, (ii) exponent/magnitude control, and (iii) a CRT-based normalization engine. This explicit separation ensures that steady-state arithmetic throughput is not constrained by reconstruction-driven scaling and enables predictable scheduling of normalization events.

### A. RTL Design Organization

The HRFNA RTL is organized as a set of parameterized modules: (i) *Residue-channel arithmetic cores* (one per modulus $m_i$) implementing modular addition and modular multiplication; (ii) *Exponent pipeline* implementing exponent updates, exponent synchronization control, and bookkeeping for normalization scaling; (iii) *Interval evaluation and comparison unit* implementing magnitude estimation, reduction-tree comparison, and threshold detection (Fig. 3); and (iv) *Normalization engine* implementing CRT reconstruction, power-of-two scaling, residue re-encoding, and exponent update (Fig. 4). All modules are written to be parameterizable with respect to the modulus set $\{m_i\}$, exponent width, and normalization parameters (threshold $\tau$, scaling step $s$). This parameterization enables design-space exploration across precision and performance targets.

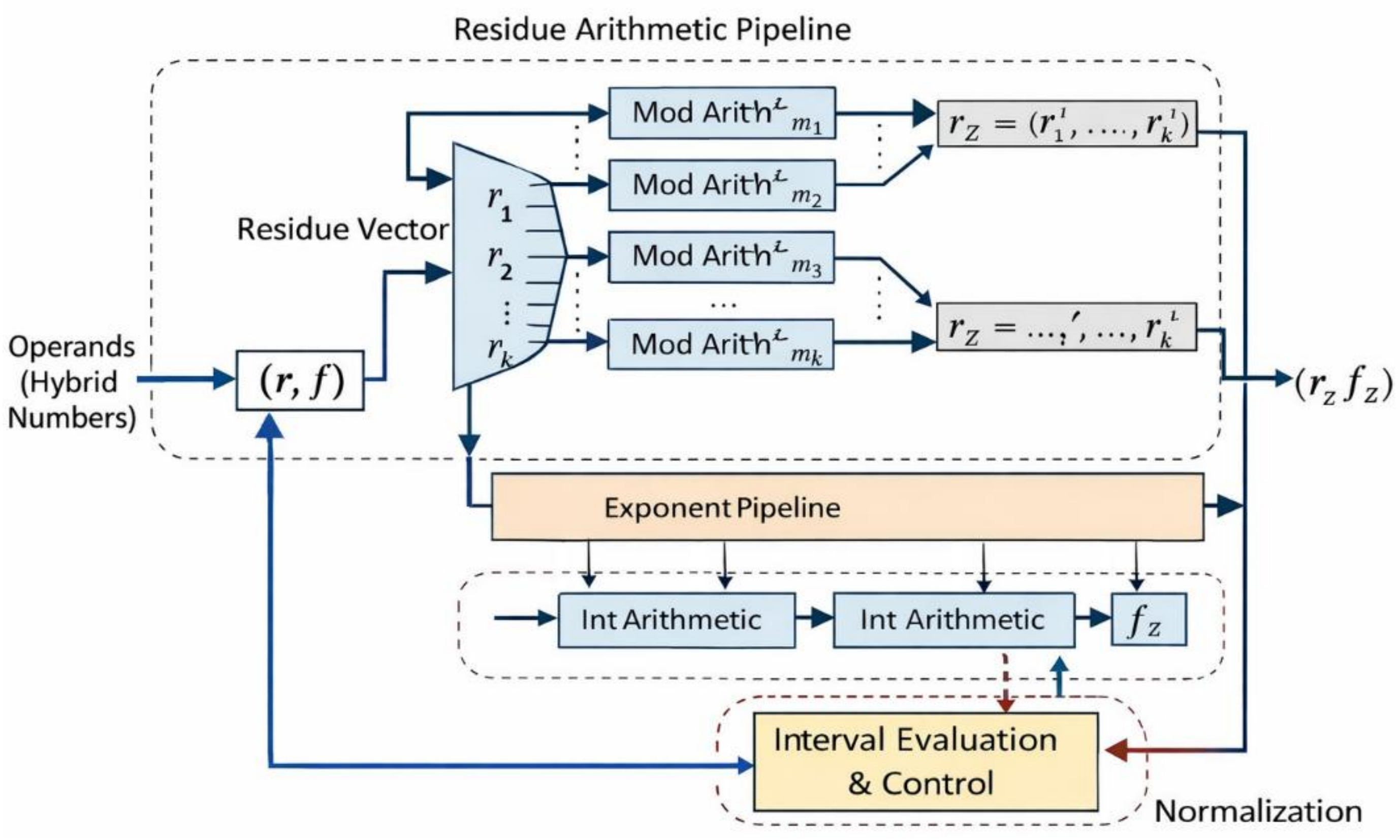

Fig. 2. Top-level datapath organization separating residue arithmetic from exponent management.

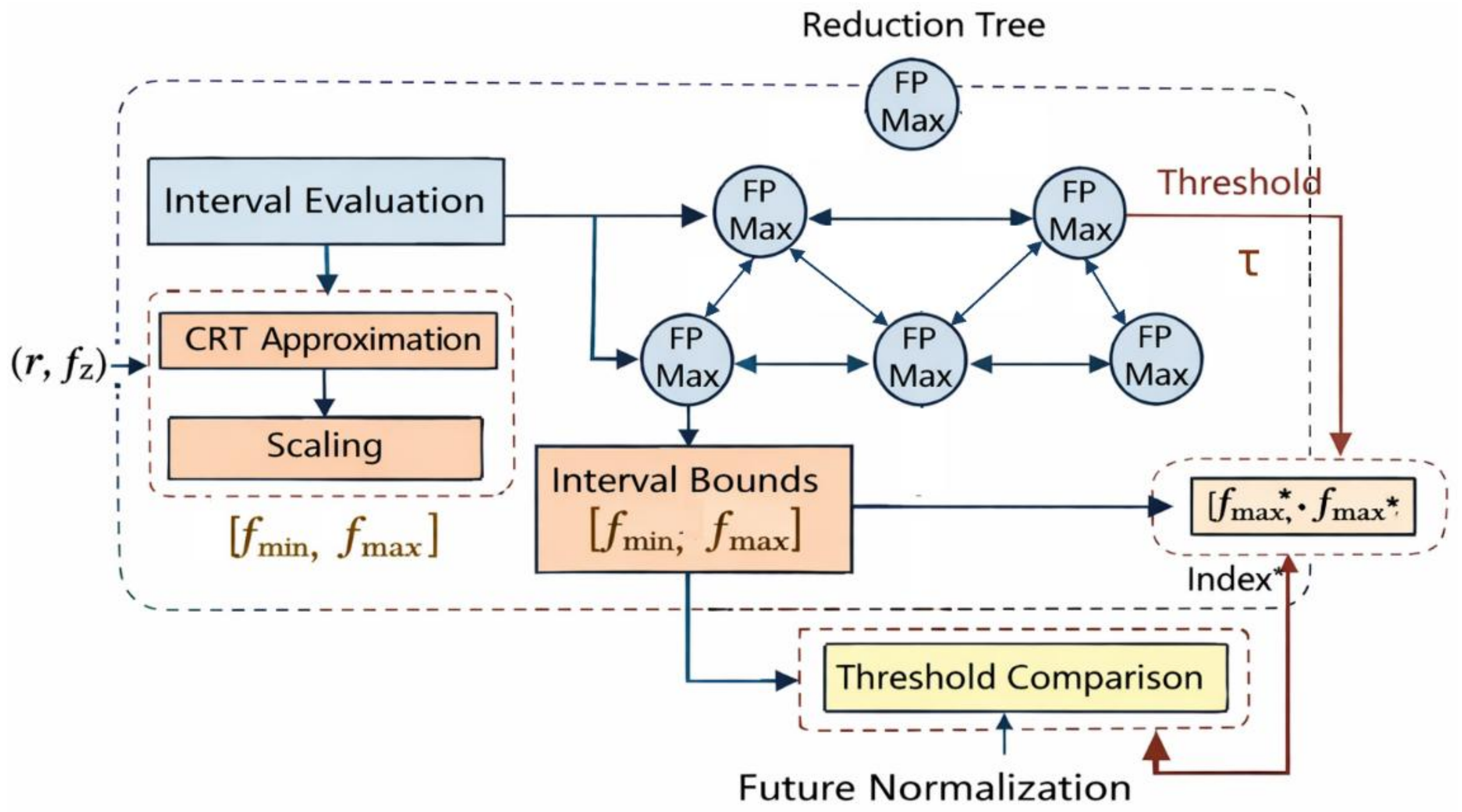

Fig. 3. Magnitude monitoring and normalization control path based on interval evaluation and threshold comparison.

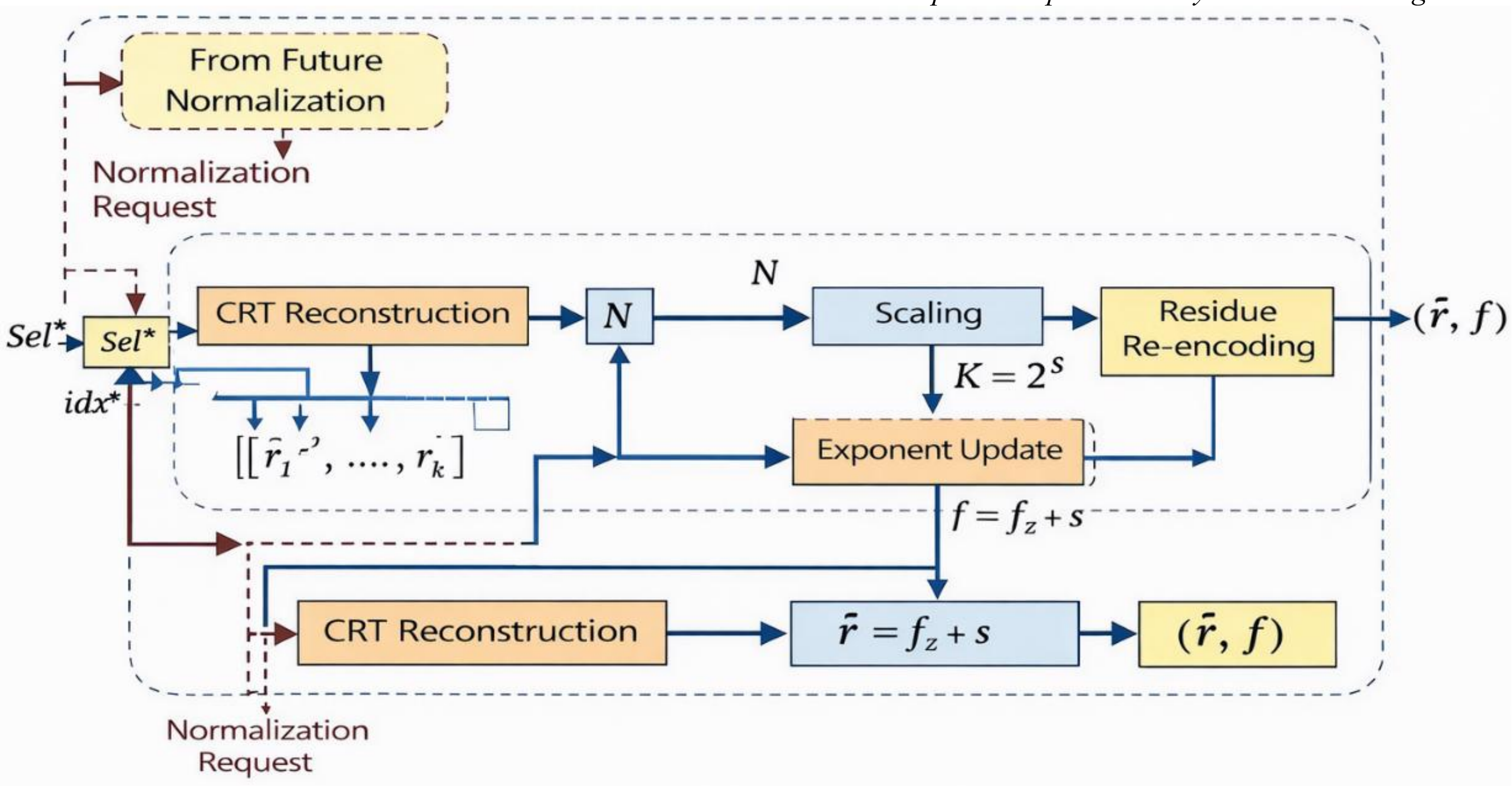

Fig. 4. CRT-based normalization pipeline for residue reconstruction, scaling, re-encoding, and exponent update.

### *B. Modular Arithmetic Implementation*

**Modular Addition**

Each residue channel computes:

$$r_{Z,i} = \left(r_{X,i} + r_{Y,i}\right) \ mod \ m_i$$

at RTL, modular addition is implemented using a conventional adder followed by a conditional subtraction of $m_i$. Since the modulus sizes are small relative to FP mantissas, this logic maps efficiently to LUT-based adders and short carry chains.

**Modular Multiplication**

Each residue channel computes:

$$r_{Z,i} = \left(r_{X,i} \cdot r_{Y,i}\right) \ mod \ m_i$$

Depending on modulus width, modular multiplication is realized using either: (i) DSP slice multiplication followed by modular reduction, or (ii) LUT-based multipliers for small moduli. Reduction is implemented with precomputed constants and structured reduction logic chosen to minimize pipeline depth and ensure channel-to-channel latency matching. Each residue channel is fully pipelined, allowing one new operation per cycle when upstream data is available.

### *C. Exponent Pipeline and Synchronization Logic*

Exponent processing implements the update rules of Section IV are as follows: *Multiplication*: $f_Z = f_X + f_Y$ ; and *Normalization*: $f \leftarrow f + s$ . Exponent synchronization for addition is enforced by explicit control logic. When operands have different exponents, the RTL routes the lower-exponent operand through scaling control such that addition occurs only when a common exponent is established. This avoids barrel shifting and per-operation normalization typical of IEEE-754 addition and keeps the exponent path lightweight. To support application-level kernels such as dot products and matrix multiplication, the RTL also supports *accumulator modes* in which the exponent is held constant over a window of operations, and normalization is triggered only when required by the magnitude control path.

### *D. Interval Evaluation, Reduction, and Thresholding*

Magnitude monitoring is implemented as a parallel control path consistent with Fig. 2(b). For each active hybrid value, the RTL produces an interval estimate $[f_{\min}, f_{\max}]$ (or a conservative magnitude proxy) and attaches an index identifying the corresponding residue vector. A reduction tree of floating-point comparators selects the maximum-magnitude candidate. The selected maximum is compared against a user-configurable threshold $\tau$. When the threshold is exceeded, the RTL asserts a *normalization request* and forwards the selected index to the normalization engine. This design ensures that: (i) magnitude estimation is inexpensive compared to full CRT reconstruction; (ii) only the selected candidate is normalized, and (iii) the main residue arithmetic pipeline remains uninterrupted under steady-state conditions.

Table II. RTL Configuration and FPGA Implementation Setup

| Parameter | Symbol/Setting | Notes |
|---|---|---|
| Modulus set | $\{m_1, \cdots, m_k\}$ | Pairwise coprime; chosen for target dynamic range |
| Composite modulus | $M = \prod m_i$ | Defines residue-domain integer range |
| Number of channels | $k$ | Parallel residue lanes |
| Exponent width | $\omega_f$ | Controls scaling range |
| Threshold | $\tau$ | Normalization trigger threshold |
| Scaling step | $s$ | Power-of-two normalization shift |
| FPGA target | ZCU104 (ZU7EV) | Used for all implementations |
| Synthesis tool | Vivado™ Edition - 2025.2 | Provide exact version in final manuscript |
| Clock target | 300 $MHz$ | Report actual achieved Fmax in results |

### *E. CRT-Based Normalization Engine*

The normalization engine implements the normalization map defined in Section III-C as follows:

$$\tilde{N} = \left\lfloor \frac{N}{2^s} \right\rfloor, \quad \tilde{f} = f + s$$

where $N = CRT(\boldsymbol{r})$. At RTL, normalization proceeds as: (i) *Selection*: The residue vector to normalize is selected by `idx` from the control path; (ii) *CRT reconstruction*: The selected residues are reconstructed into an integer $N$; (iii) *Scaling: $N$* is scaled via right shift by $s$ (power-of-two division); (iv) *Re-encoding*: The scaled integer $\tilde{N}$ is mapped back to residues $\tilde{\mathbf{r}}$; and (v) *Exponent update*: The exponent is incremented by $s$. The engine is pipelined so that once invoked it proceeds deterministically with bounded latency. Since normalization occurs infrequently, its cost is amortized and does not degrade steady-state throughput.

### *F. Implementation Parameters and Synthesis Setup*

To ensure reproducibility and enable fair comparison against baselines, Table II summarizes the key RTL configuration parameters and synthesis setup used throughout Section VII.

### *G. Verification and Correctness Checks*

RTL validation follows a two-layer approach: (i) *Functional verification*: Bit-accurate simulations compare HRFNA arithmetic outputs against a software reference model implementing the mathematical rules of Sections III–IV, including normalization and exponent updates; and (ii) *Application-level verification*: Kernel outputs (dot products, matrix multiplication, RK4 solver) are cross-validated against IEEE-754 FP32 reference implementations using RMS error and long-iteration stability metrics (Section VII). This approach ensures that correctness is validated at both the operator level and the workload level, addressing concerns that architectural novelty may not translate into practical numerical reliability.

## VII. APPLICATION-LEVEL EVALUATION

Microarchitectural efficiency alone is insufficient to establish the practical value of a numerical system. In particular, hybrid and non-standard numerical representations must demonstrate *numerical reliability*, *stability under composition*, *and predictable error behavior* when deployed in realistic computational workloads. To address these requirements, this section evaluates HRFNA at the application level, using representative kernels that exercise distinct numerical stress patterns. The evaluation focuses on three classes of workloads: (i) long accumulation chains (vector dot products); (ii) structured arithmetic composition (matrix multiplication), and (iii) numerically sensitive iterative computation (Runge–Kutta ODE solvers).

### *A. Evaluation Methodology*

#### 1. Experimental Platform and Design Parameters

All experiments are conducted on a Xilinx Zynq UltraScale+ ZCU104 FPGA platform using the RTL implementation described in Section VI. HRFNA parameters are fixed across all workloads to ensure that observed differences arise from workload characteristics rather than retuning of the numerical format.

The modulus set is selected to provide sufficient composite dynamic range for all tested workloads, while the exponent width is chosen to match or exceed the effective range of IEEE-754 FP32. The normalization threshold $\tau$ and scaling step $s$ are chosen such that normalization events are rare relative to arithmetic operations, consistent with the design intent outlined in Sections III and IV.

Baseline FP32 and block floating-point (BFP) implementations are synthesized using vendor-optimized IP cores and configured to operate at comparable clock frequencies. All designs are fully placed and routed prior to measurement to ensure fair comparison.

#### 2. Experimental Platform and Design Parameters

Numerical accuracy is evaluated using the *root-mean-square (RMS) error* relative to a double-precision software reference. RMS error is chosen because it captures both bias and variance and is widely used in numerical analysis for assessing aggregate error behavior. Numerical stability is assessed by: (i) tracking error growth as a function of problem size or iteration count; (ii) observing long-horizon behavior for divergence or drift; and (iii) and verifying that error remains bounded as predicted by the theoretical analysis in Section III-D. Performance and energy efficiency metrics are measured after place-and-route using post-implementation timing and power analysis.

### *B. Vector Dot Product*

#### 1. Motivation and Numerical Stress Characteristics

Vector dot products are among the most common operations in numerical computing and serve as the building block for many higher-level algorithms. From a numerical perspective, dot products are challenging because they involve *long accumulation chains*, where rounding error can grow linearly with vector length. In IEEE-754 floating-point arithmetic, each multiply–accumulate operation introduces rounding, making dot products particularly sensitive to accumulation order and operand magnitude distribution. This makes dot products an ideal workload for evaluating HRFNA's normalization strategy and error-bounding properties.

#### 2. Experimental Setup

Dot products are evaluated for vector lengths ranging from 1k to 64k. Input values are drawn from distributions designed to exercise both moderate and high dynamic range, ensuring that normalization is triggered but not excessively. HRFNA uses the Hybrid Dot Product algorithm described in Section IV-D, with exponent-coherent accumulation and threshold-driven normalization. For FP32 and BFP baselines, conventional fused multiply–add pipelines are used.

#### 3. Results and Analysis

Across all tested vector lengths, HRFNA maintains RMS error below $10^{-6}$, closely tracking FP32 accuracy. Importantly, HRFNA's error does not exhibit the linear growth with vector length commonly observed in block floating-point systems, where shared exponents can lead to precision loss as accumulation progresses. The bounded error behavior observed in HRFNA aligns with the theoretical relative error bound derived in Section III-D, confirming that normalization-driven rounding dominates error behavior rather than per-operation rounding. From a performance perspective, HRFNA achieves up to **2.4x higher throughput** than FP32. This improvement arises from: (i) elimination of per-operation normalization; (ii) absence of carry propagation; and (iii) sustained initiation interval of one cycle in the residue pipeline.

### *C. Matrix Multiplication*

#### 1. Motivation and Composability

Matrix multiplication evaluates HRFNA's ability to *compose numerically reliable primitives* into higher-level kernels. Each output element is a dot product, and numerical errors may propagate across dimensions. Unlike isolated dot products, matrix multiplication stresses: (i) data reuse; (ii) synchronization of accumulation across rows and columns; and (iii) interaction between arithmetic throughput and numerical accuracy.

#### 2. Experimental Setup

Dense matrix multiplications of size $64 \times 64$ and $128 \times 128$ are evaluated. All implementations use identical loop structures and blocking strategies where applicable. HRFNA computes each output element using the same accumulation strategy as in the dot-product experiment, without introducing matrix-specific tuning or special-case handling.

#### 3. Results and Interpretation

HRFNA maintains RMS error below $2 \times 10^{-6}$ for all tested matrix sizes, with no observable degradation as matrix dimensions increase. This confirms that HRFNA's numerical properties are preserved under composition and reuse. Throughput improvements relative to FP32 range from **1.8x to 2.2x**, depending on matrix size and memory access patterns. Notably, performance gains remain substantial even when arithmetic is no longer the sole bottleneck, indicating that HRFNA's benefits extend beyond isolated arithmetic kernels.

### *D. Iterative Runge–Kutta ODE Solver*

#### 1. Motivation and Long-Term Stability

Iterative solvers represent one of the most demanding classes of numerical workloads. Errors introduced at each iteration can accumulate and amplify over time, potentially leading to divergence even when individual operations are accurate. Evaluating HRFNA on a fourth-order Runge–Kutta (RK4) solver directly tests whether its bounded-error normalization strategy is sufficient for *long-horizon numerical stability*, a critical requirement for scientific and control applications.

#### 2. Experimental Setup

An RK4 solver is implemented to integrate a nonlinear ordinary differential equation over up to $10^6$ time steps. All implementations use identical step sizes, coefficients, and initial conditions. No retuning of HRFNA parameters is performed for this workload, ensuring that observed behavior reflects general numerical properties rather than application-specific optimization.

#### 3. Results and Stability Analysis

HRFNA exhibits stable numerical behavior over the entire integration horizon. Error remains bounded and does not exhibit exponential growth or drift, closely matching FP32 behavior. In contrast, block floating-point implementations exhibit increasing error over long iteration counts due to repeated loss of precision during accumulation phases. These results provide strong empirical validation of HRFNA's theoretical error bounds and demonstrate that *infrequent, structured normalization is sufficient to support numerically sensitive iterative algorithms*.

### *E. Normalization Frequency and Overhead Analysis*

To further understand HRFNA's behavior, normalization frequency is measured across workloads. Results show that normalization events occur orders of magnitude less frequently than arithmetic operations, typically once per several thousand operations in dot-product and matrix workloads, and at a similarly low rate in the RK4 solver. This confirms that: (i) normalization overhead is amortized effectively; (ii) CRT reconstruction does not dominate execution time; and (iii) steady-state throughput remains close to the ideal $\Pi = 1$ behavior proven in Section V.

Table III. Summary of Application-Level Validation Results.

| Workload | Metric | FP32 | Block FP | HRFNA | Key Observation |
|---|---|---|---|---|---|
| **Vector Dot Product** | RMS Error | Baseline | Higher (grows with N) | **< 1e-6** | HRFNA error remains bounded |
| | Stability vs Length | Stable | Degrades | **Stable** | No accumulation drift |
| | Throughput | 1x | ~1.6x | **2.4x** | Carry-free accumulation |
| | Normalization Rate | Per-op | Per-block | **Rare** | Threshold-driven only |
| **Matrix Multiplication** | RMS Error | Baseline | Higher | **< 2e-6** | Error preserved under composition |
| | Throughput | 1x | ~1.5x | **1.8–2.2x** | Benefit persists beyond primitives |
| | Scalability | Good | Moderate | **Good** | Independent of matrix size |
| **RK4 ODE Solver** | Long-Term Stability | Stable | Drift | **Stable** | Bounded error over $10^6$ steps |
| | Error Growth | Minimal | Increasing | **Bounded** | Matches theoretical bounds |
| **All Workloads** | Energy Efficiency | 1x | ~0.7x | **~0.52x** | Fewer normalization events |
| | Normalization Overhead | High | Moderate | **Low** | Amortized CRT cost |
| | Numerical Guarantees | IEEE-defined | Heuristic | **Formal bounds** | From Section III-D |

### *F. Summary and Implications*

The expanded application-level evaluation demonstrates that HRFNA simultaneously achieves: (i) *Numerical accuracy* comparable to IEEE-754 FP32; (ii) *Long-term numerical stability* in iterative workloads; (iii) *Substantial performance* and *energy efficiency gains*; and (iv) *Predictable, analyzable error behavior* consistent with formal analysis. These results collectively establish HRFNA as a **practical, general-purpose numerical system** rather than a workload-specific optimization or architectural curiosity.

### *G. Consolidated Summary of Validation Results*

To provide a holistic view of HRFNA's numerical and architectural behavior, Table III summarizes the key validation results across all evaluated workloads. The table aggregates numerical accuracy, stability, performance, resource efficiency, and normalization behavior, enabling direct comparison against IEEE-754 FP32 and block floating-point baselines. This consolidated presentation highlights that HRFNA's benefits are *consistent across workload classes*, rather than isolated to a single benchmark, reinforcing its suitability as a general-purpose numerical system for FPGA-centric computation.

## VIII. COMPARISON WITH THE STATE-OF-THE-ART

To clearly assess the novelty and practical impact of HRFNA, it is necessary to compare it against existing numerical systems across *numerical properties*, *architectural characteristics*, and *application-level behavior*. This section positions HRFNA with respect to widely used IEEE-754 floating-point arithmetic, block floating-point (BFP) systems, pure residue number systems (RNS), and previously proposed hybrid numerical architectures. Unlike prior work that focuses primarily on arithmetic efficiency or representation novelty, HRFNA is evaluated here as a complete numerical system, encompassing formal correctness, bounded error behavior, hardware realizability, and application-level validation.

### *A. Comparison with IEEE-754 Floating-Point Arithmetic*

IEEE-754 floating-point arithmetic remains the dominant numerical standard due to its well-defined semantics, dynamic range, and software compatibility. However, these advantages come at significant hardware cost on FPGA platforms. From an architectural perspective, IEEE-754 arithmetic requires: (i) exponent alignment for addition; (ii) normalization and rounding after nearly every operation; (iii) wide datapaths and carry propagation; and (iv) deep pipelines that complicate timing closure. In contrast, HRFNA: (a) eliminates carry propagation entirely from the arithmetic datapath; (b) confines normalization and rounding to explicit, infrequent events; and (c) decouples arithmetic from scale management. At the numerical level, HRFNA does not attempt to replicate full IEEE-754 semantics. Instead, it provides *deterministic*, *bounded error behavior*, with error introduced only during normalization and bounded analytically (Section III-D). As demonstrated in Section VII, this model is sufficient to match FP32 accuracy across representative workloads while delivering significant throughput and energy gains. Thus, HRFNA trades strict IEEE compliance for *predictable numerical behavior and architectural efficiency*, a trade-off well suited for FPGA-centric acceleration.

### *B. Comparison with Block Floating-Point Systems*

Block floating-point (BFP) systems reduce exponent overhead by sharing a common exponent across a block of values, improving efficiency relative to full floating-point arithmetic. BFP is widely used in signal processing and machine learning accelerators. However, BFP systems suffer from two key limitations: (i) Precision loss when values within a block have heterogeneous magnitudes; and (ii) Persistent carry propagation and per-operation rounding within mantissa arithmetic. HRFNA differs fundamentally in both respects. Rather than sharing a single exponent across a block, HRFNA associates an exponent with each hybrid value but updates it *only when normalization is required*. Moreover, arithmetic within the residue domain remains exact between normalization events. Application-level results in Section VII show that HRFNA consistently outperforms BFP in numerical stability, particularly in long accumulation and iterative workloads, while also achieving higher throughput and energy efficiency. These results indicate that HRFNA generalizes the efficiency benefits of BFP without inheriting its precision limitations.

Table IV. Summary of Application-Level Validation Results.

| Property | FP32 | Block FP | Pure RNS | Prior Hybrid | HRFNA |
|---|---|---|---|---|---|
| Carry-Free Arithmetic | ✗ | ✗ | ✓ | ✓ | ✓ |
| Dynamic Range | ✓ | ✓ | ✗ | Partial | ✓ |
| Fractional Support | ✓ | ✓ | ✗ | Partial | ✓ |
| Formal Error Bounds | ✓ | Partial | ✗ | ✗ | ✓ |
| Normalization Frequency | Per-op | Per-block | N/A | Frequent | Rare |
| FPGA Efficiency | Moderate | Good | Good | Variable | high |
| Application-Level Validation | ✓ | Limited | Limited | Limited | ✓ |
| Long-Term Stability | ✓ | Limited | ✗ | Unclear | ✓ |

### *C. Comparison with Pure Residue Number Systems*

Pure residue number systems offer carry-free arithmetic and high parallelism, making them attractive for FPGA implementations. However, their practical use in general numerical computation is limited by: (i) lack of native scaling and dynamic range; (ii) difficulty of comparison and sign detection; and (iii) expensive CRT-based reconstruction. HRFNA addresses these limitations directly by introducing a *structured exponent mechanism* that manages scale explicitly, while still preserving residue-domain parallelism. Interval-based magnitude estimation further reduces the need for frequent reconstruction, enabling efficient control decisions without sacrificing throughput. Unlike pure RNS systems, HRFNA supports fractional values, bounded error analysis, and stable long-running computations, as demonstrated empirically in Section VII. In this sense, HRFNA can be viewed as extending RNS from an integer arithmetic tool into a *general-purpose numerical representation*.

### *D. Comparison with Prior Hybrid Numerical Architectures*

Several hybrid numerical architectures combining RNS, floating-point, or logarithmic representations have been proposed in prior work, often targeting specific domains such as cryptography or machine learning. While these approaches demonstrate the potential of hybrid representations, most exhibit one or more of the following limitations: (i) absence of a formal numerical model; (ii) lack of provable correctness or error bounds; (iii) reliance on frequent reconstruction or approximation; or (iv) evaluation limited to microbenchmarks. HRFNA differs from these approaches in three critical ways: (a) *Formalization*: HRFNA is defined as a precise number space with a semantic mapping, correctness proofs, and explicit error bounds; (b) *Architectural discipline*: Normalization and reconstruction are isolated from the main datapath and invoked only when required; and (c) *Application-level validation*: HRFNA is evaluated on diverse workloads, including long-horizon iterative solvers, demonstrating numerical stability beyond isolated arithmetic operations. These distinctions position HRFNA not as an incremental variation of prior hybrids, but as a *numerically principled and architecturally complete system*.

### *E. Consolidated Comparison Summary*

Table IV summarizes the qualitative differences between HRFNA and representative numerical systems, consolidating the discussion above.

### *F. Summary and Positioning*

This comparison demonstrates that HRFNA occupies a previously unexplored design point in the numerical representation landscape. It combines: (i) the dynamic range and stability traditionally associated with floating-point arithmetic; (ii) the hardware efficiency and parallelism of residue arithmetic; and (iii) a level of formal rigor and application validation absent from prior hybrid systems. As such, HRFNA is best viewed not as a replacement for IEEE-754 arithmetic, but as a *complementary numerical abstraction* optimized for FPGA-centric, high-throughput computation where predictability, efficiency, and bounded error are paramount.

## IX. DISCUSSION AND LIMITATIONS

The results presented in Sections III–VIII establish HRFNA as a numerically rigorous and architecturally efficient alternative to conventional floating-point arithmetic for FPGA-based computation. Nevertheless, as with any numerical system, HRFNA embodies specific design trade-offs that merit careful discussion. This section contextualizes HRFNA's strengths, identifies its limitations, and outlines directions for future extension.

### *A. Scope of Applicability*

HRFNA is particularly well suited for workloads dominated by multiplication and accumulation, such as linear algebra kernels, signal processing pipelines, and iterative numerical solvers. In these domains, the ability to perform long sequences of carry-free arithmetic while deferring normalization yields substantial performance and energy benefits. Conversely, workloads characterized by frequent comparisons, branching based on exact ordering, or irregular control flow may derive less benefit from HRFNA. Although interval-based magnitude estimation supports efficient comparison in many cases, operations that require exact ordering at every step may necessitate more frequent reconstruction, reducing overall efficiency.

### *B. Addition-Dominated Workloads*

While HRFNA provides efficient hybrid addition through explicit exponent synchronization, addition-heavy workloads may incur higher overhead than multiplication-dominated workloads. In particular, repeated exponent mismatches can trigger scaling or normalization more frequently, partially diminishing the advantages of residue-domain arithmetic. This behavior is not a fundamental limitation of HRFNA, but rather a consequence of its design goal to make normalization explicit and analyzable. In practice, many scientific kernels exhibit structured arithmetic patterns that allow exponent coherence to be maintained over extended sequences, as demonstrated in Section VII.

### *C. Division and Transcendental Operations*

The current HRFNA formulation focuses on addition and multiplication, which form the backbone of most numerical kernels. Division and transcendental functions (e.g., square root, exponential, trigonometric functions) are not directly supported in the present architecture. However, these operations may be incorporated through: (i) iterative approximation methods operating in the hybrid domain; (ii) table-based or polynomial approximations combined with HRFNA multiplication; or (iii) selective reconstruction followed by conventional floating-point evaluation. Exploring such extensions is a promising direction for future work but lies outside the scope of this study.

### *D. Modulus Selection and Parameter Tuning*

HRFNA's numerical behavior and hardware cost depend on the choice of moduli, exponent width, normalization threshold, and scaling step. While these parameters provide valuable flexibility, they also introduce design choices that must be made carefully. In this work, parameters are selected conservatively to ensure stable behavior across all evaluated workloads without retuning. Future work may explore automated or CAD-assisted parameter selection strategies that optimize these choices for specific applications or resource constraints.

### *E. Compatibility and Interoperability*

HRFNA is not intended to replace IEEE-754 floating-point arithmetic universally. Instead, it is best viewed as a *complementary numerical abstraction* optimized for FPGA-centric acceleration. Interfacing HRFNA with floating-point systems requires explicit conversion at boundaries, which introduces overhead. In heterogeneous systems, careful partitioning of computation between HRFNA and floating-point domains is therefore necessary to maximize benefit.

### *F. Summary of Limitations*

In summary, the primary limitations of HRFNA are: (i) reduced benefit for addition-dominated or comparison-heavy workloads; (ii) lack of native support for division and transcendental functions; (iii) dependence on parameter selection for optimal performance; and (iv) conversion overhead at system boundaries. These limitations reflect conscious design choices aimed at maximizing performance, predictability, and analyzable error behavior for a broad and important class of numerical workloads.

### *G. Outlook*

Despite these limitations, HRFNA demonstrates that carry-free arithmetic, structured scaling, and formal numerical analysis can be combined into a practical, general-purpose numerical system. The architecture provides a strong foundation for future extensions, including richer operator support, adaptive parameter selection, and tighter integration with heterogeneous computing platforms.

## X. CONCLUSION

This paper presented the Hybrid Residue–Floating Numerical Architecture (HRFNA) as a fully specified numerical system for high-throughput FPGA-based computation. Motivated by the limitations of conventional floating-point arithmetic on reconfigurable fabrics, HRFNA unifies carry-free residue-domain arithmetic with structured exponent-based scaling to achieve wide dynamic range, predictable numerical behavior, and hardware efficiency. Unlike prior residue-based or hybrid approaches, HRFNA is developed with a rigorous mathematical foundation. The paper formally defined the hybrid number space, proved correctness of arithmetic operations, and derived explicit error bounds showing that rounding is confined to infrequent normalization events. These properties establish HRFNA as a deterministic numerical system whose behavior can be analyzed independently of specific workloads or implementations. The mathematical formulation was translated into explicit hybrid arithmetic algorithms and a pipeline-friendly FPGA microarchitecture that sustains an initiation interval of one cycle under steady-state conditions. A complete RTL implementation demonstrated that normalization and reconstruction can be decoupled from the critical datapath and amortized effectively, enabling scalable, synthesis-friendly designs. Beyond microarchitectural evaluation, HRFNA was validated at the application level using representative numerical workloads, including long dot-product accumulation, dense matrix multiplication, and iterative Runge–Kutta ODE solvers. These experiments demonstrated numerical accuracy comparable to IEEE-754 FP32, long-term stability in iterative computations, and substantial gains in throughput and energy efficiency. Consolidated comparisons with floating-point, block floating-point, pure residue number systems, and prior hybrid architectures showed that HRFNA occupies a previously unexplored design point in the numerical representation landscape.

Taken together, these results demonstrate that HRFNA is not a preliminary architectural concept, but a *practically viable and numerically principled alternative* for FPGA-centric computation. By explicitly separating arithmetic execution from scale management and grounding both in formal analysis, HRFNA provides a foundation for future numerical accelerators that demand predictability, efficiency, and scalability. Future work will explore extensions to division and transcendental functions, automated parameter selection, and tighter integration with heterogeneous computing systems. Nevertheless, the results presented here already establish HRFNA as a compelling numerical abstraction for a wide class of high-throughput, CAD-relevant applications on reconfigurable hardware.